\documentclass[10pt,pra,aps,twocolumn,showpacs,amsmath,amssymb, floatfix]{revtex4-1}
\usepackage{times}
\usepackage{textcomp}
\usepackage{graphicx}
\usepackage{bm}
\usepackage{braket}
\usepackage[breaklinks]{hyperref}
\hypersetup{colorlinks=true, linkcolor=blue, citecolor=blue, filecolor=blue, urlcolor=blue}
\usepackage{qcircuit}

\begin{document}

\title{Entanglement phase transition with spin glass criticality}

\author{Jeremy C\^{o}t\'{e}}
\author{Stefanos Kourtis}

\affiliation{D\'{e}partement de physique \& Institut quantique, Universit\'{e} de Sherbrooke, Sherbrooke, Qu\'{e}bec, J1K 2R1, Canada}

\date{\rm\today}

\begin{abstract}

We define an ensemble of random Clifford quantum circuits whose output state undergoes an entanglement phase transition between two volume-law phases as a function of measurement rate. Our setup maps exactly the output state to the ground space of a spin glass model. We identify the entanglement phases using an order parameter that is accessible on a quantum chip. We locate the transition point and evaluate a critical exponent, revealing spin glass criticality. Our work establishes an exact statistical mechanics theory of an entanglement phase transition.

\end{abstract}

\maketitle

%\section{Introduction}

Generating entanglement between the qubits in a quantum processor is one of the prerequisites for useful quantum computation. If the entanglement entropy of an ensemble of quantum states only scales polylogarithmically with the number of qubits, we can simulate the states on a classical computer in polynomial time~\cite{vidal_efficient_2003, van_den_nest_classical_2007, brandao_area_2013}. In contrast, it is unlikely we can classically simulate quantum algorithms which involve generic states with larger amounts of entanglement~\cite{jozsa_role_2003, brandao_area_2013}. If there is any potential for quantum speedups, it will be in this regime.

In the circuit model of quantum computation, a quantum algorithm is a time sequence of unitary operations on the initial state of a qubit array, followed by a measurement of the output state. While multi-qubit unitary operations increase entanglement on average, decoherence events---often modeled as mid-circuit projective measurements---decrease it. This raises a question: How high a rate of decoherence will eliminate the potential for quantum computational speedup?

Random quantum circuit ensembles with variable mid-circuit measurement rates offer a minimal setup to investigate this question~\cite{li_measurement-driven_2019, skinner_measurement-induced_2019, gullans_scalable_2020, bao_theory_2020, zabalo_critical_2020, potter_entanglement_2021}. These circuits include randomly chosen one- and two-qubit unitary gates, which generally increase the entanglement entropy. Then, projective measurements simulate decoherence and decrease entanglement. When the rate of measurement is low, the entanglement in the system grows linearly in the system size (a volume law). By contrast, a high rate of measurement prevents entanglement from accumulating in the system, resulting in entanglement which does not grow with system size (an area law). By varying the measurement rate, a variety of random quantum circuit ensembles go through a measurement-induced phase transition.

Researchers model these phase transitions and their criticality using statistical mechanics systems.
These include percolation~\cite{li_measurement-driven_2019, skinner_measurement-induced_2019,  zabalo_critical_2020, vijay_measurement-driven_2020, jian_measurement-induced_2020, bao_theory_2020, lavasani_measurement-induced_2021, lunt_measurement-induced_2021}, quantum Ising models~\cite{block_measurement-induced_2021}, and classical Ising and Potts models~\cite{bao_theory_2020}. These mappings to statistical mechanics models only work in specific limits, such as infinite physical Hilbert space dimension~\cite{bao_theory_2020, jian_measurement-induced_2020}. Though this modeling brings theoretical insight into the transition, it is at the expense of experimental relevance of the model.

In this Letter, we introduce a random Clifford quantum circuit ensemble with a finite measurement rate, whose output state maps \emph{exactly} to the ground state manifold of a classical $p$-spin model. This ensemble of circuits undergoes a volume-to-volume-law entanglement phase transition as a function of measurement rate. We establish a correspondence between the entanglement entropy of the quantum state at the output and the ground-state entropy of the $p$-spin model. Finite-size scaling reveals spin glass criticality of the entanglement transition in the ensemble. We also calculate an order parameter which is accessible in current quantum processors. Our setup allows for experimental realization of a spin glass phase transition on a quantum chip.

\begin{figure}[t]
    \centering
    \includegraphics[width=\columnwidth]{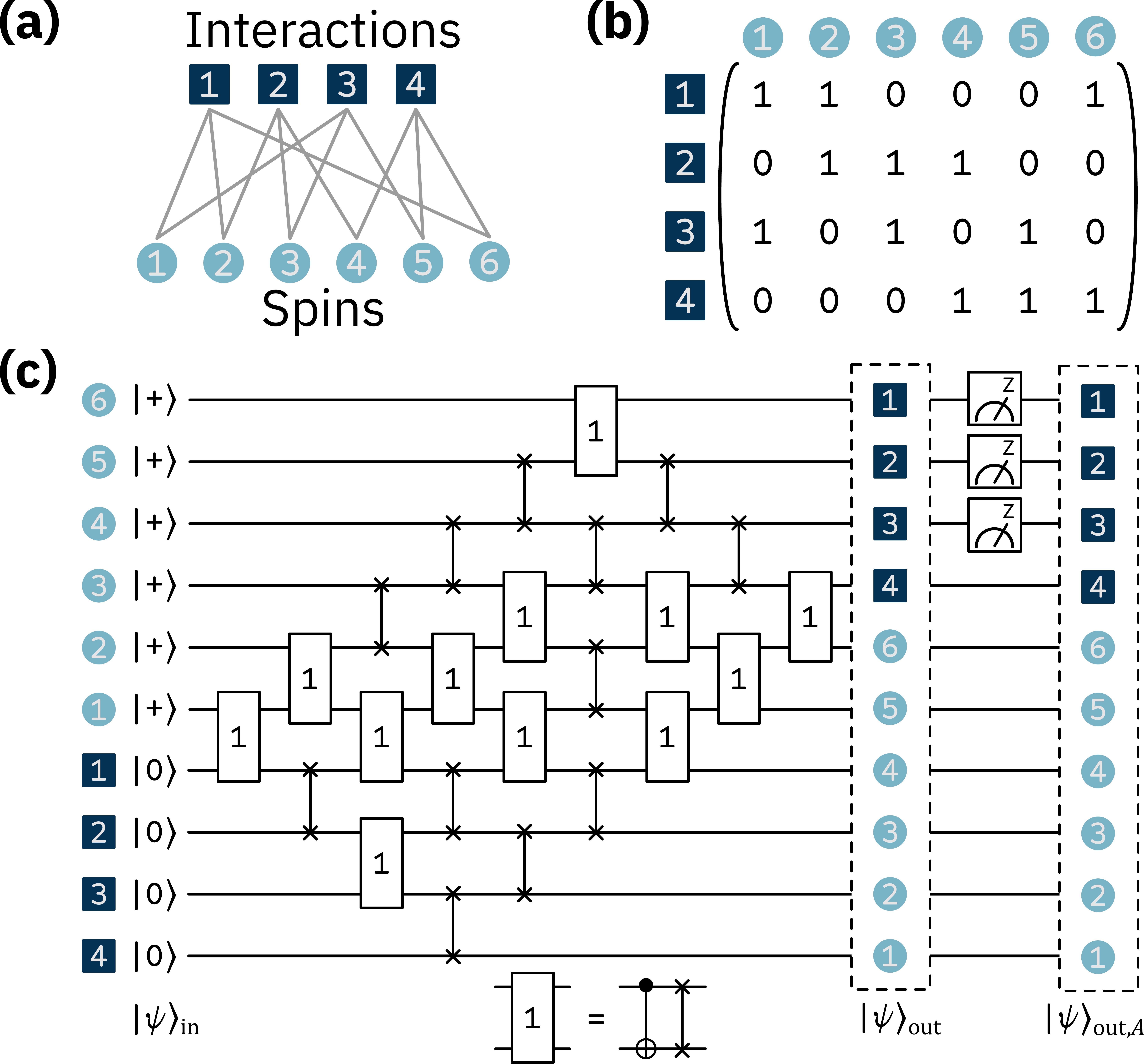}
    \caption{The spin-glass model in its various forms. (a) A graph of spins and interactions, as in Eq.~\eqref{eq:Hamiltonian}. (b) The matrix $B$ of the graph, as defined in Eq.~\eqref{eq:MatrixEquation}, where each column is a spin and each row is an interaction. (c) The quantum circuit built from $B$. A one corresponds to the rectangular gate shown in the inset, composed of a CNOT then a SWAP, whereas a zero is a SWAP gate. The initial quantum state is $\ket{\psi}_{\mathrm{in}}$, the quantum state after the applying the gates in $B$ is $\ket{\psi}_{\mathrm{out}}$ in Eq.~\eqref{eq:AlphaState}, and after measuring a region $A$ of the interaction qubits, the output state is $\ket{\psi}_{\mathrm{out}, A}$ in Eq.~\eqref{eq:MeasuredState}. The labels for the qubits in the quantum circuit specify the spins and interactions.}
    \label{fig:Model}
\end{figure}

We begin by describing the statistical mechanics system that will model our circuit ensemble. We consider a system of $N$ classical Ising spins interacting according to a bipartite graph $G = \left(V_J, V_\sigma, E \right)$ which we draw in Fig.~\ref{fig:Model}(a). There are two vertex sets: $V_J$ for interactions, and $V_\sigma$ for spins, with $\lvert V_J \rvert = M$ and $\lvert V_\sigma \rvert = N$. The edge set $E$ connects spins to interactions. Each interaction vertex $a \in V_J$ connects to three spin vertices. The Hamiltonian for a given graph $G$ is:
\begin{equation}
    \label{eq:Hamiltonian}
    H(G, \vec{\sigma}, \vec{J}) = \frac{1}{2} \sum_{a \in V_J} \left( 1 - J_a \sigma_{a_1} \sigma_{a_2} \sigma_{a_3} \right),
\end{equation}
where $a_1, a_2, a_3 \in V_\sigma$ refer to the three spin vertices connected to $a$, $J_a = \pm 1$ are the couplings, and the spins take values $\sigma_{i} = \pm 1$. We build each instance of the Hamiltonian by choosing $M$ triplets $(a_1,a_2,a_3)$ uniformly at random, with two restrictions: three distinct spin vertices per triplet and no repeated triplets. We also choose the couplings $J_a$ such that $H(G, \vec{\sigma}, \vec{J}) = 0$ for at least one spin configuration. With these choices, the model~\eqref{eq:Hamiltonian} is the unfrustrated $p$-spin model for $p=3$~\cite{franz_ferromagnet_2001, ricci-tersenghi_simplest_2001, mezard_two_2003}.

The ground states of this model have zero energy. Using the mapping $J_a = (-1)^{y_a}$ and $\sigma_{i} = (-1)^{x_{i}}$, finding the ground states is equivalent to solving the system
\begin{equation}
    \label{eq:MatrixEquation}
    B \vec{x} = \vec{y} \mod{2} \,.
\end{equation}
$B$ is a $M \times N$ binary matrix, called the biadjacency matrix of the graph $G$, whose rows represent interaction terms and columns represent spins, as shown in Fig~\ref{fig:Model}(a)-(b). If $B_{ij} = 1$, then interaction $i$ involves spin $j$. Otherwise, the entry is zero.

The number of ground states is
\begin{equation}
    \label{eq:SolutionCount}
    \mathcal{N}_{\text{GS}}(B) = 2^{N - \text{rank}(B)},
\end{equation}
where $\text{rank}(B)$ is the binary rank of $B$ and $N-\text{rank}(B)$ is the number of independent rows of $B$. The ground-state entropy is defined as
\begin{equation}
    \label{eq:GroundStateEntropy}
    S_{\text{GS}}(B) \equiv \log \mathcal{N}_{\text{GS}}(B),
\end{equation}
where we take the natural logarithm for the rest of the Letter.

For a given $B$, any unfrustrated choice of couplings in~\eqref{eq:Hamiltonian} gives the same number of ground states, since $\text{rank}(B)$ is independent of the couplings. 

We are now ready to introduce our random circuit ensemble, shown schematically in Fig.~\ref{fig:Model}(c). We use $N + M$ qubits to represent a spin system with $N$ spins and $M$ interactions. For each spin participating in an interaction, the corresponding spin qubit toggles the state of an interaction qubit via a controlled NOT (CNOT) gate. SWAP gates allow any spin qubit to toggle any interaction qubit, such that any graph $G$ can be realized in this setup. At the circuit's output, each interaction qubit stores the parity of the configuration of the three spin qubits that connect to it. Note that because we can implement the circuit described so far using only CNOT gates, this is a classical circuit. Each circuit requires $O(NM)$ gates.

To generate entanglement at the output of the circuits, we initialize the input in a superposition. Each spin qubit is put into the $\ket{+}$ state and each interaction qubit in $\ket{0}$, yielding the input state $\ket{\psi}_{\mathrm{in}} = \ket{+}^{\otimes N} \ket{0}^{\otimes M}$. After applying the gates defined by $B$, the state at the output of the circuit (before any measurements) is
\begin{equation}
    \label{eq:AlphaState}
    \ket{\psi}_{\mathrm{out}} = \frac{1}{\sqrt{2^{\text{rank}(B)}}} \sum_{\vec{y}} \ket{ \vec{y} } \ket{\{ \vec{x} \, : \,  B\vec{x} = \vec{y} \}},
\end{equation}
where $\ket{\{ \vec{x} \, : \,  B\vec{x} = \vec{y} \}}$ is an equal superposition of $\mathcal{N}_{\mathrm{GS}}$ solutions to Eq.~\eqref{eq:MatrixEquation} for a given choice of $B$ and $\vec{y}$. The normalization of Eq.~\eqref{eq:AlphaState} comes from the number of vectors $\ket{\vec{y}}$ in the sum, which is $2^{\text{rank}(B)}$. To see this, note that the unitary that describes a circuit of CNOT and SWAP gates is a permutation matrix. This means that if we start with a superposition of $2^N$ states at the input, we necessarily end up with a superposition of $2^N$ states at the output. But Eq.~\eqref{eq:SolutionCount} dictates that for each $\ket{\vec{y}}$ we must have $2^{N-\text{rank}(B)}$ configurations, and so there must be $2^{\text{rank}(B)}$ terms in the sum.

If we measure the interaction qubits at the output of the circuit, then the measurement outcome is a parity assignment $\vec{y}_{\mathrm{out}}$. The remaining quantum state over the spin qubits is a superposition of ground states to Hamiltonian~\eqref{eq:Hamiltonian} for the choice of couplings $\vec{J} = (-1)^{\vec{y}_{\mathrm{out}}}$. Subsequently measuring the spin qubits picks out one ground state $\vec{\sigma} = (-1)^{\vec{x}_{\mathrm{out}}}$.

Now, suppose we measure only a subset $A$ consisting of the first $k \le M$ interaction qubits and get outcome $\vec{y}_{\text{out}} = \left( y_{\mathrm{out},1}, \ldots, y_{\mathrm{out},k}  \right)$. The updated state is:
\begin{equation}
    \label{eq:UpdateRule}
    \ket{\psi}_{\mathrm{out},A} =  \frac{\ket{\vec{y}_{\text{out}}}\bra{\vec{y}_{\text{out}}} \otimes I}{\sqrt{\Pr (\vec{y}_{\text{out}})}} \ket{\psi}_{\mathrm{out}},
\end{equation}
where $I$ is the identity operator on the remaining $N-k$ qubits and $\Pr (\vec{y}_{\text{out}})$ is the probability of getting outcome $\vec{y}_{\text{out}}$. Define $B_A$ as the submatrix of $B$ consisting of the rows corresponding to interactions in $A$. There are $2^{\text{rank}(B_A)}$ possible vectors $\vec{y}_{\text{out}}$ which have solutions to $B_A \vec{x} = \vec{y}_{\text{out}}$, so $\Pr(\vec{y}_{\text{out}}) = 2^{-\text{rank}(B_A)}$. The post-measurement state is then
\begin{equation}
    \label{eq:MeasuredState}
    \ket{\psi}_{\mathrm{out},A} = \frac{1}{\sqrt{2^{\text{rank}(B)-\text{rank}(B_A)}}}  \sum_{ \{ \vec{y} \,:\, \vec{y}_k = \vec{y}_{\mathrm{out}} \} } \ket{\vec{y}} \ket{\{ \vec{x} \, : \,  B\vec{x} = \vec{y} \}} \,,
\end{equation}
where $\vec{y}_k$ is the vector made from the first $k$ components of $\vec{y}$.

Equation~\eqref{eq:MeasuredState} is a Schmidt decomposition between the interaction qubits and the spin qubits, with the normalization being the Schmidt coefficients $\left\{ \lambda_i \right\}$. The states $\ket{\vec{y}}$ are orthonormal (each $\ket{\vec{y}}$ is unique with unit length), and so are the normalized states $\ket{\{ \vec{x} \, : \,  B\vec{x} = \vec{y} \}}$, since all the components of each state are solutions to Eq.~\eqref{eq:MatrixEquation} for only one $\ket{\vec{y}}$.

The entanglement entropy between the interaction qubits and the spin qubits in the post-measurement state is
\begin{equation}
    \label{eq:EntanglementEntropy}
    S ( \ket{\psi}_{\mathrm{out},A} ) \equiv - \sum_{i} \lambda_i^2 \log{\lambda_i^2} = \Big( \text{rank}(B)-\text{rank}(B_A) \Big) \log{2}.
\end{equation}
In the following, we measure all entropies in units of $\log2$. We also focus only on instances with sufficiently large $M \ge N$ such that $\text{rank}(B) = N$ with high probability. The entanglement entropy is then
\begin{equation}
    \label{eq:EntropyMap}
    S ( \ket{\psi}_{\mathrm{out},A} ) = N -\text{rank}(B_A) = S_{\text{GS}}(B_A),
\end{equation}
where $S_{\text{GS}}(B_A)$ is the ground-state entropy of the instance of Hamiltonian~\eqref{eq:Hamiltonian} with interaction graph $G_A$ and its associated matrix $B_A$. Equation~\eqref{eq:EntropyMap} is our main result.

Equation~\eqref{eq:EntropyMap} establishes an exact correspondence between the entanglement entropy of the post-measurement state in our random quantum circuit ensemble and the ground-state entropy of a spin-glass model. Note that while we can also calculate the entanglement entropy in Eqs.~\eqref{eq:EntanglementEntropy} and ~\eqref{eq:EntropyMap} using the stabilizer formalism~\cite{hamma_ground_2005, hamma_bipartite_2005, nahum_quantum_2017}, our result gives Eq.~\eqref{eq:EntropyMap} a physical meaning. The parameter controlling the entropy in both models is $ \alpha \equiv |A|/N$. In the spin glass model, $\alpha$ is the number of interactions per spin. In the random quantum circuit model, $\alpha$ is the number of measured interaction qubits to spin qubits (a measurement rate). This differs from other work on random quantum circuits, where the measurement rate refers to the probability of measuring any qubit during each layer of the circuit.

We can go further and obtain an exact expression for the ensemble-averaged entanglement entropy by exploiting the correspondence to the spin glass. The mean ground-state entropy density of the unfrustrated $p$-spin model at a given $\alpha$ is given by~\cite{monasson_course_2007}
\begin{equation}
    \label{eq:DensityOfSolutions}
    s_{\text{GS}}(\alpha) = \max_{0 \le d \le 1} \Big[ \left( 1 - d \right) \left( 1 - \log{\left(1 - d \right)} \right) - \alpha \left( 1 - d^3 \right) \Big],
\end{equation}
where $s_{\mathrm{GS}}(\alpha) = \langle S_{\mathrm{GS}} \rangle / N$, with $\langle \ldots \rangle$ denoting the ensemble average. This translates to the mean entanglement entropy of our random circuit ensemble via Eq.~\eqref{eq:EntropyMap}.

The ground space of Hamiltonian~\eqref{eq:Hamiltonian} exhibits a sharp transition~\cite{franz_ferromagnet_2001, ricci-tersenghi_simplest_2001} as a function of $\alpha$. The transition occurs at $\alpha_c \approx 0.918$ and coincides with the slowing down of all local relaxation dynamics. The transition causes a non-analyticity of Eq.~\eqref{eq:DensityOfSolutions} at $\alpha_c$. This transition is physically the same as the SAT-UNSAT transition in $p$-XORSAT \cite{creignou_smooth_2003, mezard_two_2003}, where the probability of finding at least one solution to the model jumps from one to zero.

For $\alpha < \alpha_c$, the system is paramagnetic. In terms of Eq.~\eqref{eq:MatrixEquation}, each interaction is approximately independent of the others, dividing the ground space in half. Here, $\text{rank}(B_A) = |A| =  N\alpha$ and the maximum in Eq.~\eqref{eq:DensityOfSolutions} occurs for $d = 0$. In the quantum circuits, measuring an interaction qubit eliminates half the ground states. The mean entanglement entropy after measuring $A$ interaction qubits then follows the volume law $\langle S \rangle \sim N \left( 1 - \alpha \right)$.

For $\alpha \ge \alpha_c$, the system is a spin glass. The overlap between ground states to Eq.~\eqref{eq:MatrixEquation} for a given $\vec{y}_{\text{out}}$ goes from zero to a finite value as $\alpha$ increases past $\alpha_c$, meaning most ground states share the same spin value on a given site~\cite{franz_ferromagnet_2001, ricci-tersenghi_simplest_2001}. Equation~\eqref{eq:GroundStateEntropy} reflects this by having the maximum occur at $d > 0$. Now, $\text{rank}(B_A) \ne N\alpha$ but the entanglement entropy still follows a volume law, just with a different slope compared to the paramagnet. This implies the ensemble of states $\left\{ \ket{\psi}_{\text{out,A}} \right\}$ undergoes an entanglement phase transition at $\alpha_c$.

To detect the criticality of our ensemble of quantum circuits, we introduce a measure we call \emph{entanglement susceptibility}. This quantity monitors the change in entanglement entropy upon measuring one more interaction qubit in the system. The idea is that, by Eqs.~\eqref{eq:EntropyMap} and~\eqref{eq:DensityOfSolutions}, the entanglement entropy is non-analytic at $\alpha_c$, then its derivative will be discontinuous.

For a given matrix $B_A$, we define the entanglement susceptibility $\chi_S$ as:
\begin{equation}
    \begin{aligned}
    \label{eq:EntanglementSusceptibility}
    \chi_S (\ket{\psi}_{\mathrm{out},A}) &\equiv - \Delta_{h} \left[ S \right] ( \ket{\psi}_{\mathrm{out},A} ) \\
    &= \Big( S ( \ket{\psi}_{\mathrm{out},A} ) - S ( \ket{\psi}_{\mathrm{out},A+h} ) \Big) \\
    &= \Big( \text{rank}(B_{A+h})-\text{rank}(B_{A}) \Big),
    \end{aligned}
\end{equation}
where $\Delta_h$ is a finite difference with a step of one extra measured interaction qubit $h$.

Because we define $\chi_S$ in terms of a difference of ranks between matrices which only differ by one row, $\chi_S \in \left\{0, 1 \right\}$. A new interaction is independent from the previous ones when $\chi_S = 1$, and dependent when $\chi_S = 0$. Each matrix $B$ produces a set of values $\left\{ \chi_S \right\}$ for each $\alpha$, and averaging over the ensemble of matrices produces the finite-size scaling curves we use to detect the transition.

As $N \rightarrow \infty$, $\langle S \rangle / N$ approaches $s_{\mathrm{GS}}(\alpha)$ in Eq.~\eqref{eq:DensityOfSolutions}. This means the entanglement susceptibility will go from a finite difference to a derivative~\cite{monasson_course_2007}:
\begin{equation}
    \label{eq:ContinuumSusceptibility}
    \lim_{N \rightarrow \infty} \langle \chi_S \rangle = - \frac{\mathrm{d} s_{\mathrm{GS}}(\alpha)}{\mathrm{d} \alpha}.
\end{equation}
Equation~\eqref{eq:ContinuumSusceptibility} provides an exact expression for the discontinuous phase transition occurring in the equation for $S_{\mathrm{GS}}(\alpha)$ in the thermodynamic limit.

We then do finite-size scaling with the entanglement susceptibility using the scaling form~\cite{Supplement, kawashima_critical_1993}
\begin{equation}
    \label{eq:FiniteSizeScaling}
    \langle \chi_{S} \rangle = f\left(\left(\alpha - \alpha_c  \right) N^{1/\nu} \right),
\end{equation}
with $\nu$ being the scaling dimension of the entanglement susceptibility. We find a value of $\alpha_{c, \, \text{scaling}} = 0.9175 \pm 0.0006$ and $\nu = 1.8 \pm 0.1$. The value of $\alpha_{c, \, \text{scaling}}$ agrees with the value in the literature of $\alpha_c \approx 0.9179$. The critical exponent $\nu$ also agrees with numerical~\cite{ricci-tersenghi_simplest_2001} and analytical~\cite{wilson_critical_2002} work on the criticality of the SAT-UNSAT transition, while being distinct from the critical exponent of $\nu \approx 1.3$ commonly found in other ensembles of random Clifford circuits~\cite{potter_entanglement_2021}.

\begin{figure}[t]
    \centering
    \includegraphics[width=\columnwidth]{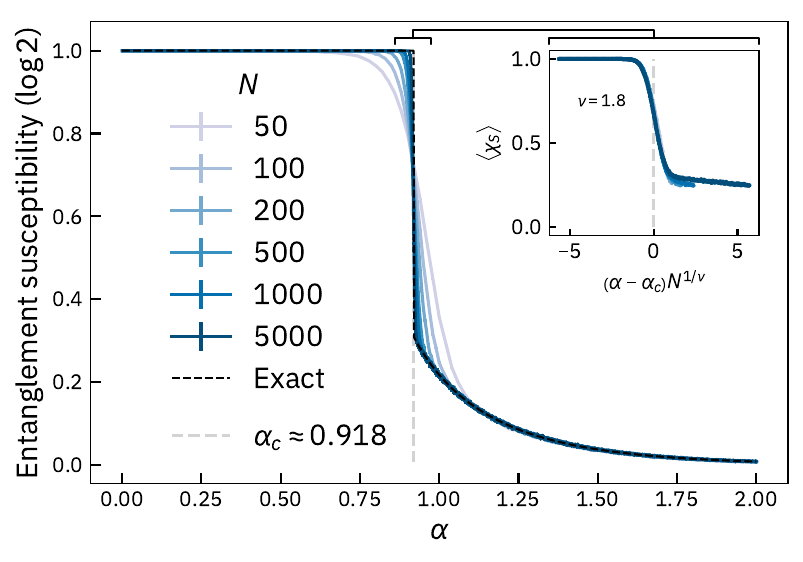}
    \caption{Finite-size scaling of the averaged entanglement susceptibility $\langle \chi_S \rangle$, in units of $\log{2}$. We average over 20,000 samples for each system size, and error bars indicate standard error of the mean. We calculate the exact expression by taking a finite derivative of Eq.~\eqref{eq:DensityOfSolutions}. The inset gives the collapsed version of the main plot within the region $\alpha_c \pm 0.05$.}
    \label{fig:insetEntanglementSusceptibility}
\end{figure}

This choice of $\alpha_c$ and $\nu$ collapses the data around the critical point as shown in Fig.~\ref{fig:insetEntanglementSusceptibility}. The transition sharpens with system size, and becomes discontinuous in the $N \rightarrow \infty$ limit, which we mark with the dashed line. We conclude that our ensemble of random quantum circuits has the same criticality as the $3$-spin model and associated Boolean satisifiablity problems.

We can also derive an order parameter for the entanglement phases in our quantum circuit model via the 3-spin model~\eqref{eq:Hamiltonian}. The order parameter for this model is the overlap $q$ of ground states~\cite{ricci-tersenghi_simplest_2001}. It is defined as~\cite{sherrington_solvable_1975}
\begin{equation}
    \label{eq:Overlap}
    q = \frac{1}{N} \sum_{i = 1}^N \langle \sigma_i \rangle^2,
\end{equation}
where $\langle \ldots \rangle$ is an average over all the $\mathcal{N}_{\mathrm{GS}}$ ground states, and $\sigma_i$ is the spin at a specific site $i$. We square the average because if not the sum over the spins in Eq.~\eqref{eq:Overlap} is the magnetization, which is on average zero for this ensemble (and spin glasses in general). We then average $q$ over many instances. The overlap captures the degree to which ground states to a given instance of Eq.~\eqref{eq:Hamiltonian} are the same.

\begin{figure}[ht]
    \centering
    \includegraphics[width=\columnwidth]{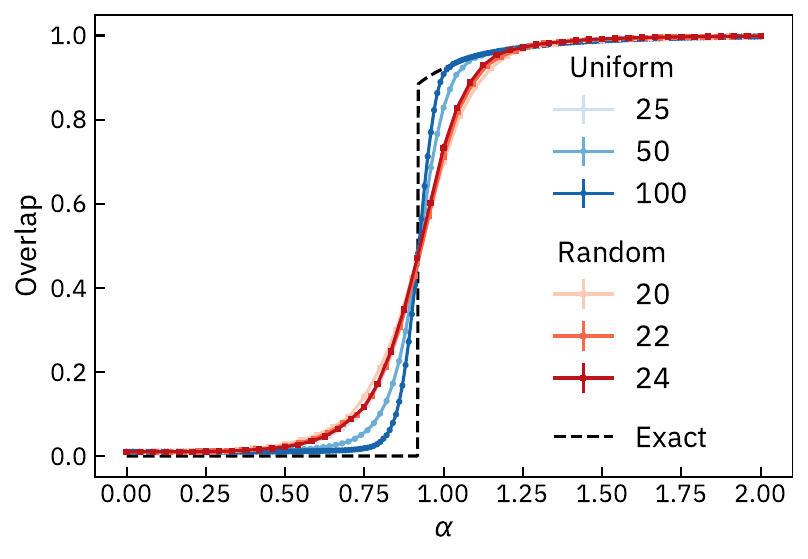}
    \caption{Overlap $q(\alpha)$ of the ground states. We average over $1000$ samples per system size. We sample $\text{min}\left(100, \mathcal{N}_{\mathrm{GS}} \right)$ ground states, and use the unique ones to calculate the overlap in Eq.~\eqref{eq:Overlap}. Error bars indicate standard error of the mean. We use either the $\ket{\psi}_{\mathrm{in}} = \ket{+}^{\otimes N} \ket{0}^{\otimes M}$ input state (blue circles), or a random input state (red squares). The dashed line is the exact overlap in the thermodynamic limit~\cite{monasson_course_2007}.}
    \label{fig:overlap}
\end{figure}

For small $\alpha$, the overlap is $q = 0$, because ground states are independent. This is true all the way until $\alpha_c$, where ground states then tend to overlap on most sites. The overlap $q$ takes on a finite value, trending to $q = 1$ as $\alpha \rightarrow \infty$. We can see this discontinuous jump in Fig.~\ref{fig:overlap}, which sharpens with system size. Figure~\ref{fig:overlap} also confirms that the overlap order parameter is accurately approximated using a small sample of the ground states.

The quantum circuits defined so far are a subset of Clifford circuits, meaning the state at any given time is a stabilizer state. Are stabilizer states a necessary condition to observe the spin-glass entanglement criticality? Figure~\ref{fig:overlap} presents evidence that the answer is ``no'' and that this critical phenomenon is more general. Instead of the initial state $\ket{\psi}_{\mathrm{in}} = \ket{+}^{\otimes N} \ket{0}^{\otimes M}$, we input the partially scrambled state $\ket{\tilde\psi}_{\mathrm{in}} = \ket{\psi_{\text{scrambled}}}^{\otimes N} \ket{0}^{\otimes M}$ into our circuits. That is, instead of putting the spin qubits in an equal superposition of all computational basis states, we put them in a random one. We then follow the same measurement protocol as for the Clifford case and compute $q(\alpha)$~\cite{Supplement}. The red curves in Fig.~\ref{fig:overlap} show that the output state still transitions at $\alpha_c$, at least for small system sizes.

The order parameter $q(\alpha)$ for our Clifford ensemble can be measured on existing quantum hardware. This requires multiple measurements $\ket{\vec{x}}$ for a given $\ket{\vec{y}}$. Note that measuring the interaction qubits at the output of the quantum circuit will in general give different vectors $\vec{y}$, requiring many shots (on the order of $2^{\mathrm{rank}(B)}$) to get two measurements results which have the same $\vec{y}$. This, however, is not necessary. Suppose we measure the pair $( \vec{x}, \vec{y} )$. We can then use Gaussian elimination to solve for a \emph{reference} configuration $\vec{x}'$ for a fixed $\vec{y}$, such that $B \vec{x}' = \vec{y}$. The reference configuration could be, for example, the configuration $\vec{x}'$ whose bit sequence represents the smallest nonzero integer. Then, observe that
\begin{equation}
    \label{eq:RotatingSolutions}
    \vec{z} \equiv \vec{x} + \vec{x}' \rightarrow B\vec{z} = B \left( \vec{x} + \vec{x}' \right) = \vec{y} + \vec{y} = \vec{0} \mod{2}.
\end{equation}
This means we can create a pair $( \vec{z}, \vec{0} )$ for any measurement of the interaction vector $\vec{y}$ we get. Because we can do Gaussian elimination in polynomial time with the system size, this means we can efficiently transform any measurement output to the $\vec{y} = \vec{0}$ interaction vector, and then use Eq.~\eqref{eq:Overlap}. This means that we can efficiently measure the order parameter $q(\alpha)$ on a quantum chip.

We note two distinctions between our setup and previous work on entanglement phase transitions in random quantum circuits. First, previous studies required a finite measurement rate throughout their circuits. For depth-$d$ circuits with $n$ qubits, this leads to $O(nd)$ measurements. In contrast, in our ensemble, measurements are only required at the output and are $O(n)$. This makes our setup simpler and may also reduce the effect of readout errors in practice. Second, our circuits only use CNOT gates (excluding the input) to encode the $p$-spin system. The ``random'' part of our ensemble is in the placement of these gates, as opposed to randomly sampling from the full Clifford gate set like in previous work. Despite these differences, our work captures the salient features of entanglement phase transitions, with the added advantage that the criticality in this case is exactly tractable.

In summary, by starting with a statistical mechanics system and embedding it into an ensemble of random Clifford circuits, we showed how our random quantum circuits display an entanglement phase transition with spin glass criticality. This quantum circuit ensemble is a physically relevant model, with connections to spin glasses and Boolean satisfiability problems. The numerical evidence we presented indicates the presence of the same transition in more general circuit classes. Our work emphasizes the feasibility of detecting this phase transition on both current and future quantum processors. Our hope is to not only use quantum processors to study phase transitions in random quantum circuits, but as experimental platforms for studying spin glass physics.

\begin{acknowledgements}
This work was supported by the Minist\`{e}re de l'\'{E}conomie et de l'Innovation du Qu\'{e}bec via its contributions to its Research Chair in Quantum Computing, and a Natural Sciences and Engineering Research Council of Canada Discovery grant. Jeremy acknowledges the support of a B2X scholarship from the Fonds de recherche--Nature et technologies and a scholarship from the Natural Sciences and Engineering Research Council of Canada [funding reference number: 456431992]. We acknowledge Calcul Qu\'{e}bec and Compute Canada for computing resources.

Jeremy performed all simulations and data analysis, made the figures, and wrote the Letter. Stefanos conceived the idea for the project, provided guidance along the way, and wrote the Letter.
\end{acknowledgements}

%merlin.mbs apsrev4-1.bst 2010-07-25 4.21a (PWD, AO, DPC) hacked
%Control: key (0)
%Control: author (72) initials jnrlst
%Control: editor formatted (1) identically to author
%Control: production of article title (-1) disabled
%Control: page (0) single
%Control: year (1) truncated
%Control: production of eprint (0) enabled
%

%%%%%%%%%% Supplementary Materials %%%%%%%%%%
\onecolumngrid
\clearpage
\begin{center}
\textbf{\large Supplemental Materials: Entanglement phase transition with spin glass criticality}
\end{center}
\makeatletter

\section{Exhaustive enumeration}
\label{sec:Exhaustive}

To do the exhaustive enumeration for the random quantum input states, we begin by defining a random complex unit vector $\ket{\psi}$ of length $2^N$, and its corresponding probability vector $\vec{P}$, with
\begin{equation}
    \label{eq:Probability}
    P_{\vec{x}} \equiv \lvert \braket{\vec{x} | \psi} \rvert ^2,
\end{equation}
where $\ket{\vec{x}}$ is the basis state $\vec{x}$ for our quantum state, and corresponds to one of the $2^N$ possible ground states we begin with.

For each interaction $a$, we calculate the probability that the state has an even or odd parity for $a$:
\begin{equation}
    \begin{aligned}
    \label{eq:InteractionProbability}
    P_{\mathrm{even}}(a) &\equiv \sum_{\vec{x}\, : \, B_a \vec{x} = 0} P_{\vec{x}}, \\
    P_{\mathrm{odd}}(a)  &\equiv \sum_{\vec{x}\, : \, B_a \vec{x} = 1} P_{\vec{x}},
    \end{aligned}
\end{equation}
where $B_a$ is the submatrix of $B$ with only the row defined by interaction $a$. Note the matrix equation indexing both sums uses binary arithmetic.

With $P_{\mathrm{even}}$ and $P_{\mathrm{odd}}$, we flip a biased coin with these probabilities to decide if we eliminate the ground states that are even or odd under interaction $a$. We eliminate the ground states by setting their probabilities to zero. We then renormalize our probability vector using the usual rule in quantum theory:
\begin{equation}
    \label{eq:RenormalizeProbability}
    P_{\vec{x}} \leftarrow \frac{P_{\vec{x}}}{P_{\mathrm{even/odd}}}.
\end{equation}
We then sample $P_{\vec{x}}$ to get our configurations for Eq.~\eqref{eq:Overlap}. We do this for each value of $\alpha \in \left[0, 2\right]$, which produces the red curves in Fig.~\ref{fig:overlap}.

\section{Finite-size scaling}
\label{sec:FiniteSizeScaling}

To get the critical exponent and threshold from the finite-size scaling, we used the same procedure of Refs.~\cite{kawashima_critical_1993, zabalo_critical_2020, lunt_measurement-induced_2021}. We will outline the key steps. We start with the scaling form:
\begin{equation}
    \chi_S = f\left(\left(\alpha - \alpha_c  \right) N^{1/\nu} \right).
\end{equation}
The idea is to have all of the data from Fig.~\ref{fig:insetEntanglementSusceptibility} collapse onto a similar curve $f$ for the correct choice of $\alpha_c$ and $\nu$. To do this, we minimize a cost function built from the data and associated error.

The data for this experiment comes in the form $\left( \alpha, \chi_S(\alpha), e(\alpha) \right)$, where $e(\alpha)$ is the standard error of the mean for the data point. We transform this into:
\begin{equation}
    \label{eq:TransformData}
    \left( x_i, y_i, e_i \right) = \left( \left[ \alpha - \alpha_c \right] N^{1/\nu}, \chi_S(\alpha), e(\alpha) \right).
\end{equation}
We sort these triples by their $x$-values for the upcoming computation. Then, our cost function $C(\alpha_c, \nu)$ is:
\begin{equation}
    \label{eq:CostFunction}
    C(\alpha_c, \nu) = \frac{1}{n-2} \sum_{i = 2}^{n-1} w\left( x_i, y_i, e_i \vert x_{i-1}, y_{i-1}, e_{i-1}, x_{i+1}, y_{i+1}, e_{i+1}  \right).
\end{equation}
The quantity in the summation is:
\begin{equation}
    \label{eq:w}
    w\left( x_i, y_i, e_i \vert x_{i-1}, y_{i-1}, e_{i-1}, x_{i+1}, y_{i+1}, e_{i+1}  \right) = \left( \frac{y_i - \Bar{y}}{\Delta \left(y_i - \Bar{y} \right)} \right)^2,
\end{equation}

\begin{equation}
    \label{eq:yBar}
    \Bar{y} = \frac{\left(x_{i+1} - x_i \right) y_{i-1} - \left(x_{i-1} - x_i \right) y_{i+1}}{\left(x_{i+1} - x_{i-1} \right)},
\end{equation}

\begin{equation}
    \label{eq:Uncertainty}
    \left[ \Delta \left(y_i - \Bar{y} \right) \right]^2 = e_i^2 + \left( \frac{x_{i+1} - x_{i}}{x_{i+1} - x_{i-1}} \right)^2 e_{i-1}^2 + \left( \frac{x_{i-1} - x_{i}}{x_{i+1} - x_{i-1}} \right)^2 e_{i+1}^2.
\end{equation}

The cost function $C(\alpha_c, \nu)$ measures how far off each data point $y_i$ is from the linear interpolation $\Bar{y}$ between the previous point $x_{i-1}$ and the next point $x_{i+1}$ in the sorted sequence. This is why the $x$-values are sorted first, and why we exclude the first and last data points from Eq.~\eqref{eq:CostFunction}. The uncertainty in Eq.~\eqref{eq:Uncertainty} is a weighted sum of the error of the current point $i$ as well as the error from the previous and next data points. If there are three identical $x$-values in a row, we skip over them in Eq.~\eqref{eq:CostFunction} (because Eqs.~\eqref{eq:yBar} and \eqref{eq:Uncertainty} will have a division by zero).

We plot the cost function $C(\alpha_c, \nu)$ for a grid of values. Because the finite-size scaling collapse only works well in the region near the critical point, we restricted our data for $C(\alpha_c, \nu)$ to the region $\alpha_c \pm 0.05$, indicated by the black connectors linking the main plot with the inset in Fig.~\ref{fig:insetEntanglementSusceptibility}.

We chose our regions to be $\alpha_c \in \left[0.91, 0.93\right]$, with a step size of $0.0005$, and $\nu \in \left[1.5, 2.2 \right]$, with a step size of $0.001$. We chose an interval for $\nu$ skewing towards smaller values because previous simulations found the exponent between $\nu = 1$ and $\nu = 2$. We used a smaller step size for $\alpha_c$ because the critical threshold is known to a better precision. The step size gives us the uncertainty in the minimum of $C(\alpha_c, \nu)$ based on the resolution of our search. 

\begin{figure}[t]
    \centering
    \includegraphics[width=0.7\textwidth]{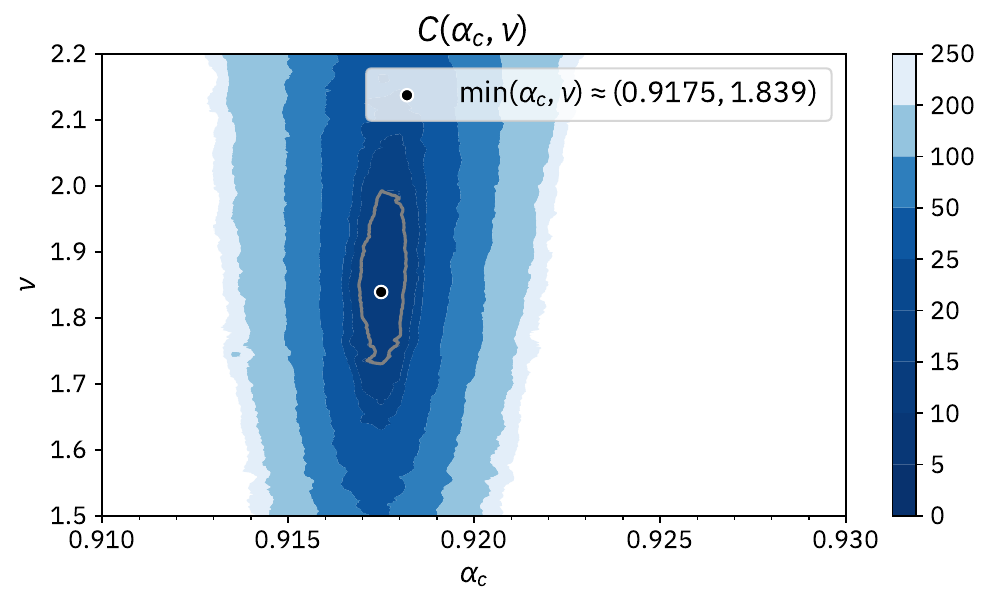}
    \caption{The cost function over a range of values of $\alpha_c$ and $\nu$. The black dot gives the minimum of the cost function, and the contour in grey is the region of uncertainty, given by $\left(1 + r\right)C_{\text{min}}(\alpha_c, \nu)$, with $r = 0.5$. Note how the minimum is in a deep basin, indicated by the scale of the colour bar. The optimal $\alpha_c$ and $\nu$ in the legend only reflect the resolution of the grid search. We used the range of $\alpha_c \pm 0.05$ for the data in Eq.~\eqref{eq:CostFunction} to compute $C(\alpha_c, \nu)$.}
    \label{fig:Landscape}
\end{figure}

We propose the following to estimate our uncertainty: We plot a contour at the level $\left(1 + r\right) C_{\text{min}}(\alpha_c, \nu)$, where $r$ is how large we look for deviations in the minimum value. For our work, we chose $r = 0.5$. Other choices are possible, and will grow or shrink the uncertainty accordingly. However, from Fig.~\ref{fig:Landscape} we can see that the minimum for the cost function resides in a deep basin, which suggests the values of $\alpha_c$ and $\nu$ we report are close to the true minimum. We take the uncertainty in $\alpha_c$ to be half the distance between the maximum and minimum values of the horizontal axis along the contour, and the uncertainty in $\nu$ to be half the distance between the maximum and minimum values of the vertical axis along the contour.

Doing so gives the following values for the critical point and critical exponent:
\begin{equation}
    \label{eq:CriticalPoints}
    \alpha_c = 0.9175 \pm 0.0006, \,\,\, \nu = 1.8 \pm 0.1.
\end{equation}
These results agree with the criticality for the SAT-UNSAT transition in Ref.~\cite{ricci-tersenghi_simplest_2001}. The authors showed the exponent $\nu$ drifts from 1 to 2 as the system size increases (top inset of Fig.~3 in that work). Analytical work~\cite{wilson_critical_2002} on the critical exponents of general satisfiability problems showed that in the $N \rightarrow \infty$ limit, the exponent should be $\nu \geq 2$. Our result is consistent with both.

Appendix A of Ref.~\cite{skinner_measurement-induced_2019} provides a second technique to find the uncertainty. Using this method (with at least three systems per iteration), we find $\alpha_c = 0.917 \pm 0.002$ and $\nu = 1.9 \pm 0.3$, which agree with analytical and numerical results on the spin glass transition~\cite{ricci-tersenghi_simplest_2001,wilson_critical_2002}.

\end{document}